\documentclass[preprintnumbers,article,amsmath,amssymb,floatfix,10pt,prd,onecolumn,superscriptaddress,nofootinbib]{revtex4-2}

\usepackage{titlesec}
\usepackage[toc]{appendix}

\titleformat*{\section}{\LARGE\bfseries}
\titleformat*{\subsection}{\Large\bfseries}
\titleformat*{\subsubsection}{\large\bfseries}
\titleformat*{\paragraph}{\large\bfseries}
\titleformat*{\subparagraph}{\large\bfseries}
\usepackage{bm}
\usepackage{amsfonts}
\usepackage{latexsym}
\usepackage[latin1]{inputenc}
\usepackage{graphicx}
\usepackage{amsmath}
\usepackage{palatino}
\usepackage{mathpazo}
\usepackage{textcomp}
\usepackage{float}
\usepackage{booktabs}
\usepackage{dcolumn}
\usepackage{ragged2e}
\usepackage{hyperref}
\hypersetup{colorlinks,citecolor=blue}
\hypersetup{colorlinks=true,linkcolor=red,filecolor=magenta,    urlcolor=blue}
\usepackage{amsthm}
\usepackage{xcolor}
\usepackage{orcidlink}
\usepackage{epsfig}
\usepackage[caption=false]{subfig}
\usepackage{commath}
\usepackage{cancel, ulem}

\usepackage{multirow}
\newcommand{\be}{\begin{equation}}
\newcommand{\ee}{\end{equation}}
\newcommand{\bea}{\begin{eqnarray}}
\newcommand{\eea}{\end{eqnarray}}
\newcommand{\eeas}{\end{eqnarray*}}
\newcommand{\beas}{\begin{eqnarray*}}
\def\jnl@style{\it}
\def\aaref@jnl#1{{\jnl@style#1}}

\def\aaref@jnl#1{{\jnl@style#1}}

\def\aj{\aaref@jnl{AJ}}                   % Astronomical Journal
\def\apj{\aaref@jnl{ApJ}}                 % Astrophysical Journal
\def\apjl{\aaref@jnl{ApJ}}                % Astrophysical Journal, Letters
\def\apjs{\aaref@jnl{ApJS}}               % Astrophysical Journal, Supplement
\def\apss{\aaref@jnl{Ap\&SS}}             % Astrophysics and Space Science
\def\aap{\aaref@jnl{A\&A}}                % Astronomy and Astrophysics
\def\aapr{\aaref@jnl{A\&A~Rev.}}          % Astronomy and Astrophysics Reviews
\def\aaps{\aaref@jnl{A\&AS}}              % Astronomy and Astrophysics, Supplement
\def\mnras{\aaref@jnl{Mon.~Not.~Roy.~Astron.~Soc.}}             % Monthly Notices of the RAS
\def\prd{\aaref@jnl{Phys.~Rev.~D}}        % Physical Review D
\def\prc{\aaref@jnl{Phys.~Rev.~C}}  % Physical Review C
\def\prl{\aaref@jnl{Phys.~Rev.~Lett.}}    % Physical Review Letters
\def\qjras{\aaref@jnl{QJRAS}}             % Quarterly Journal of the RAS
\def\skytel{\aaref@jnl{S\&T}}             % Sky and Telescope
\def\ssr{\aaref@jnl{Space~Sci.~Rev.}}     % Space Science Reviews
\def\zap{\aaref@jnl{ZAp}}                 % Zeitschrift fuer Astrophysik
\def\nat{\aaref@jnl{Nature}}              % Nature
\def\aplett{\aaref@jnl{Astrophys.~Lett.}} % Astrophysics Letters
\def\apspr{\aaref@jnl{Astrophys.~Space~Phys.~Res.}} % Astrophysics Space Physics Research
\def\physrep{\aaref@jnl{Phys.~Rep.}}      % Physics Reports
\def\physscr{\aaref@jnl{Phys.~Scr}}       % Physica Scripta
\def\commat{\aaref@jnl{Comm.~Math.~Phys.}}              % Communications in Mathematical Physics
\def\science{\aaref@jnl{Science}}               % Science
\def\cqg{\aaref@jnl{Classical Quant.~Grav.}}            % Classical and Quantum Gravity
\def\jpcs{\aaref@jnl{JPCS}}                                     % Journal of Physics Conference Series
\def\ijmpd{\aaref@jnl{Int.~J.~Mod.~Phys.~D}}                    % International Journal of Modern Physics D
\def\grg{\aaref@jnl{Gen.~Relat.~Gravit.}}               % General Relativity and Gravitation
\def\rpp{\aaref@jnl{Rep.~Prog.~Phys.}}          % Reports on Progress in Physics
\def\npa{\aaref@jnl{Nucl.~Phys.~A}}        % Nuclear Physics A
\def\lrr{\aaref@jnl{Living Rev.~Rel.}}                   % Living reviews in relativity
\def\jcap{\aaref@jnl{J.~Cosmology Astropart.~Phys.}}    % Journal of cosmology and astroparticle physics
\def\rmp{\aaref@jnl{Rev.~Mod.~Phys.}}   %Reviews of modern physics
\def\epjc{\aaref@jnl{Eur.~Phys.~J.~C}} 
\def\plb{\aaref@jnl{~Phy.~Lett.~B}} 
\def\mpla{\aaref@jnl{Mod.~Phy.~Lett.~A}} 
\def\arxiv{\aaref@jnl{arxiv.org}}

\allowdisplaybreaks[1]
\begin{document}

\title{Anisotropic Universe in $f(Q,T)$ gravity, a novel study}
\author{Tee-How Loo\orcidlink{0000-0003-4099-9843}}
\email{looth@um.edu.my}
\affiliation{Institute of Mathematical Sciences, Faculty of Science, Universiti Malaya, 50603 Kuala Lumpur, Malaysia}
\author{M. Koussour\orcidlink{0000-0002-4188-0572}}
\email{pr.mouhssine@gmail.com}
\affiliation{Quantum Physics and Magnetism Team, LPMC, Faculty of Science Ben
M'sik,\\
Casablanca Hassan II University,
Morocco.}
\author{Avik De\orcidlink{0000-0001-6475-3085}}
\email{de.math@gmail.com}
\affiliation{Department of Mathematical and Actuarial Sciences, Universiti Tunku Abdul Rahman, Jalan Sungai Long,
43000 Cheras, Malaysia}

%\date{}

\footnotetext{The research was supported by the Ministry of Higher Education (MoHE), through the Fundamental Research Grant Scheme (FRGS/1/2021/STG06/UTAR/02/1). }

\begin{abstract}
 $f(Q,T)$ theory of gravity is very recently proposed to incorporate within the action Lagrangian, the trace $T$ of the energy-momentum tensor along with the non-metricity scalar $Q$. The cosmological application of this theory in a spatially flat isotropic and homogeneous Universe is well-studied. However, our Universe is not isotropic since the Planck era and therefore to study a complete evolution of the Universe we must investigate the $f(Q,T)$ theory in a model with a small anisotropy. This motivated us to presume a locally rotationally symmetric (LRS) Bianchi-I spacetime and derive the motion equations. We analyse the model candidate $f(Q,T)=\alpha Q^{n+1}+\beta T$, and to constrain the parameter $n$, we employ the statistical Markov chain Monte Carlo (MCMC) method with the Bayesian approach using two independent observational datasets, namely, the Hubble datasets, and Type Ia supernovae (SNe Ia) datasets.   
\end{abstract}

\maketitle
%\footnotetext{   }

\section{Introduction}

Several recent Type Ia supernovae (SNe Ia) datasets \cite{Riess, Perlmutter}
and Planck Collaboration \cite{Planck2020} results have revealed that the
Universe is expanding at an accelerated rate. To explain this late-time
cosmic acceleration in the realm of general relativity (GR), several ideas
have been presented in the literature. The most prominent candidate causing
this acceleration is thought to be some yet undetected form of Dark Energy
(DE), which cannot be described by the baryonic matter. However, such a
cosmic scenario is plagued with several issues \cite{weinberg/1989},
promoting the development of alternative gravity models. The first in line
and simplest form of such modified gravity theories must be the $f(\mathring{R})$ theories, where the Ricci-scalar $\mathring{R}$ in the original Einstein-Hilbert action was replaced by an arbitrary but viable function $f(\mathring{R})$ \cite{fR}. Several curvature based modified gravity theories were proposed and analysed in the literature, for a thorough survey one can see \cite{cantata,modifiedgrav} and the references therein. In this research direction, extending the popular $f(\mathring{R})$ gravity, a matter-geometry coupling in the form of $f(\mathring{R},T)$ theory was first proposed by Harko et al \cite{harkofRT} and later frequently investigated in cosmological research \cite{fRT,fRT1}. Studies of $f(\mathring{R},T)$ gravity coupled with a real scalar field was also conducted in the inflationary paradigm \cite{fRT2}. However, the most general technique to couple a scalar field with the gravitational degrees of freedom with second-order Euler-Lagrange field equations is given by the well-known Horndeski Lagrangian, initiated in \cite{hordensk}. This theory can only incorporate the Riemannian curvature tensor and the second derivatives of the scalar field in a severely constrained form to maintain the Euler-Lagrange field equations in second order; even though the theory contains arbitrary functions of the scalar field and its kinetic component. This might be motivated to survive the Ostrogradsky instability, which was present in the higher-order theories of gravity, fourth-order $f(\mathring{R})$ theory being an exception. In particular, it is forbidden for the curvature invariants to arise freely via arbitrary functions. This is due to the fact that the curvature invariants already contain second derivatives of the metric, which, in most cases, would result in higher-order equations \cite{hor,hor1}. After the recent discovery of GW170817, most of Horndeski's terms are severely constrained by the tensor mode propagation speed \cite{gw}.

On the other hand, an equivalent formulation of gravity was proposed on a flat spacetime geometry based solely either on the torsion (TEGR) or the non-metricity (STEGR), the first is known as metric teleparallel theories and the second as symmetric teleparallel theories. Nester and Yo \cite{Nester} proposed the latter from an action term $\int{Q\sqrt{-g}d^4x}$. Due to its dependence on the dark sector, Jimenez et al. later extended it to formulate the $f(Q)$ gravity \cite{coincident} such that the late-time acceleration could be demonstrated from the additional geometric components. In STEGR, an extension was also proposed by considering a linear combination of all the possible quadratic contractions of the nonmetricity tensor \cite{23,40}. Scalar fields were also coupled to the nonmetricity scalar $Q$ in \cite{41,42}. The Hordenski type theory was very recently proposed in both metric and symmetric teleparallel geometries, respectively in \cite{hor2} and \cite{hor3} and thorough investigation is still due. Even though most of the terms are the same as in the original Hordenski theory in GR, a much richer phenomenology is noticed in the latter two counterparts.

In the recent past, a tremendous amount of works were carried out in the $f(Q)$ theories \cite{dynamical2,lcdm,accfQ1,accfQ2,accfQ3,fQfT, fQfT1, fQfT3, gde, fQfT2,
de/epjc,ad/ec, zhao, ad/bianchi,ad/ec, lin, cosmography, signa, red-shift,perturb, dynamical1}. 

Very recently, a matter-geometry coupling in the form of $f(Q,T)$ theories were proposed in which the Lagrangian was represented by a viable function of the non-metricity scalar $Q$, and the trace $T$ of the energy-momentum tensor \cite{Yixin/2019}. Harko argued that this dependence can be caused by exotic imperfect fluids or quantum phenomena \cite{Harko}. The STEGR, and specially the current $f(Q,T)$ theory is naturally a second-order theory, unlike the curvature-based original Hordenski theory which was constructed with additional constraints to be of second-order. After the first article, several works were published in this new gravity theories \cite{Arora1,Bhattacharjee,Zia,Godani}. However, all the
existing works were carried out in a background of spatially flat
homogeneous and isotropic Friedmann-Lema\^itre-Robertson-Walker (FLRW)
spacetime. Whereas, there are sufficient evidences to support a not so
symmetrical looking Universe, atleast in its beginning near the initial
singularity \cite{A2,A4,A5}. Moreover, some of the anisotropic Bianchi
models (models I, VII0, V, VIIh, and IX) can be interpreted as the
homogeneous limit of linear cosmological perturbations of the FLRW spacetime 
\cite{pb1,pb2}. The homogeneous and isotropic model naturally cannot provide
a complete account of evolution of Universe. One must relax the assumption
of FLRW geometry from the very start and investigate the transition from an
anisotropic and inhomogeneous state to the observed level of homogeneity and
isotropy. A number of recent articles in several modified gravity theories
can be cited \cite{B1, B2, B3, B4, B5, B6} for a diverse field of
investigation in the background of anisotropic Bianchi type Universe models.
In the present discussion we consider a special type of Bianchi Universe,
the locally rotationally symmetric (LRS) Bianchi type-I model to denote the
anisotropic state of the Universe, given by the metric in the Cartesian
coordinates 
\begin{equation}
ds^{2}=-dt^{2}+A^{2}(t)dx^{2}+B^{2}(t)(dy^{2}+dz^{2}).  \label{metric}
\end{equation}
Here, $A\left( t\right) $ and $B\left( t\right) =C\left( t\right) $ are the
metric potentials~that are time-dependent scale factors. All of our results
can be extended to the Bianchi type-I model without much effort.

In this paper, we analyse the model candidate $f(Q,T)=\alpha Q^{n+1}+\beta T$,
and as a special case we briefly discuss the model $f(Q,T)=\alpha Q+\beta T$, 
in the background of an anisotropic Universe with a well-known physically motivated condition of proportionality between the expansion and shear scalar. The present research focuses on observable evidence from SNe, Cosmic Microwave
Background (CMB), and Baryon Acoustic Oscillations (BAO), all of which are shown to be useful in constraining cosmological models. The Hubble
parameter $H(z)$ datasets reveal the complicated structure of the expansion
of the Universe. The ages of the most massive and slowly developing galaxies
provide direct measurements of the $H(z)$ at different redshifts $z$,
culminating in the construction of a new type of standard cosmological probe 
\cite{jim}. In this paper, we present 31 Hubble expansion observations
spread using the differential age approach \cite{Hubble}. Scolnic et al.
previously posted Pantheon, a massive SNe datasets with 1048 locations
across the redshift range $z\in \lbrack 0.01,2.3]$ \cite{Scolnic}. The
Hubble and SNe Ia are used in our analysis to constrain the cosmological
model.

The paper is structured as follows: After introduction, in Section \ref{sec2}%
, we provide an outline of $f(Q,T)$ gravity, followed by the motion
equations and some results crucial for the current study in Section \ref%
{sec3}. In Section \ref{sec4}, we propose the cosmological model used in the
paper, along with computation of certain parameters. Then we proceed to
analyze the model $f(Q,T)=\alpha Q^{n+1}+\beta T$ in Subsection \ref{subsec1}%
, and the brief account of the special linear case in Subsection \ref%
{subsec2}. In Section \ref{sec6}, we use Hubble and SNe Ia datasets to
constrain the model parameters. Finally, in Section \ref{conclusion}, we
summarise our findings.

%%%%%%%%%%%%%%%%%%%%%%%%%%%%%%%%%%%%%%%%%%%%%%%%%%%%%%%%%%%%%%%%%
%%%%%%%%%%%%%%%%%%%%%%%%%%%%%%%%%%%%%%%%%%%%%%%%%%%%%%%%%%%%%%%%%

\section{The mathematical formulation}

\label{sec2} As an extension of symmetric teleparallel gravity theory, the $%
f(Q,T)$-gravity theory is also constrained by the curvature free and torsion
free conditions, i.e., $R^{\rho }{}_{\sigma \mu \nu }=0$ and $T^{\rho
}{}_{\mu \nu }=0$. The disformation tensor is defined as the difference
between the associated connection $\Gamma ^{\lambda }{}_{\mu \nu }$ and the
Levi-Civita connection $\mathring{\Gamma}^{\lambda }{}_{\mu \nu }$ 
\begin{equation}
L^{\lambda }{}_{\mu \nu }=\Gamma ^{\lambda }{}_{\mu \nu }-\mathring{\Gamma}%
^{\lambda }{}_{\mu \nu }\,.  \label{connc}
\end{equation}%

It can also be expressed as 
\begin{equation*}
L^{\lambda }{}_{\mu \nu }=\frac{1}{2}(Q^{\lambda }{}_{\mu \nu }-Q_{\mu
}{}^{\lambda }{}_{\nu }-Q_{\nu }{}^{\lambda }{}_{\mu })\,,
\end{equation*}%
where $Q_{\lambda \mu \nu }:=\nabla _{\lambda }g_{\mu \nu }$ is the
non-metricity tensor. The non-metricity scalar $Q$ reads as \cite{zhao} 
\begin{equation}
Q=Q_{\lambda \mu \nu }P^{\lambda \mu \nu }=-\frac{1}{2}Q_{\lambda \mu \nu
}L^{\lambda \mu \nu }+\frac{1}{4}Q_{\lambda }Q^{\lambda }-\frac{1}{2}%
Q_{\lambda }\tilde{Q}^{\lambda }\,,  \label{Q}
\end{equation}%
where 
\begin{equation}
P^{\lambda }{}_{\mu \nu }:=\frac{1}{4}\left( -2L^{\lambda }{}_{\mu \nu
}+Q^{\lambda }g_{\mu \nu }-\tilde{Q}^{\lambda }g_{\mu \nu }-\frac{1}{2}%
\delta _{\mu }^{\lambda }Q_{\nu }-\frac{1}{2}\delta _{\nu }^{\lambda }Q_{\mu
}\right) \,,  \label{P}
\end{equation}%
is the superpotential tensor, $Q_{\lambda }=Q_{\lambda \mu }{}^{\mu }$ and $%
\tilde{Q}_{\lambda }=Q_{\nu \lambda }{}^{\nu }$. The action of $f(Q,T)$%
-gravity is defined as \cite{Yixin/2019} 
\begin{equation*}
S=\int \left[ \frac{1}{2\kappa }f(Q,T)+\mathcal{L}_{M}\right] \sqrt{-g}%
\,d^{4}x\,,
\end{equation*}%
where $g:=\det [g_{\mu \nu }]$, % is the determinant of the metric tensor, 
$\mathcal{L}_{M}$ is the Lagrangian for matter source and $T$ is the trace
of the stress energy tensor $T_{\mu \nu }$, which is defined as 
\begin{equation*}
T_{\mu \nu }=-\frac{2}{\sqrt{-g}}\frac{\delta (\sqrt{-g}\mathcal{L}_{M})}{%
\delta g^{\mu \nu }}\,.
\end{equation*}%

The following metric field equation can be obtained after varying the action
with respect to the metric 
\begin{equation}
\frac{2}{\sqrt{-g}}\nabla _{\lambda }(\sqrt{-g}f_{Q}P^{\lambda }{}_{\mu \nu
})-\frac{1}{2}fg_{\mu \nu }+f_{T}(T_{\mu \nu }+\Theta _{\mu \nu
})+f_{Q}(P_{\nu \rho \sigma }Q_{\mu }{}^{\rho \sigma }-2P_{\rho \sigma \mu
}Q^{\rho \sigma }{}_{\nu })=\kappa T_{\mu \nu }\,,  \label{FE1}
\end{equation}%
where $f_{Q}=f_{Q}\left( Q,T\right) $ and $f_{T}=f_{T}\left( Q,T\right) $
denote the partial derivative of $f=f\left( Q,T\right) $ with respect to $Q$
and $T$ respectively, and 
\begin{equation*}
\Theta _{\mu \nu }=\frac{g_{\alpha \beta }\delta T_{\alpha \beta }}{\delta
g^{\mu \nu }}\,.
\end{equation*}%

Noticing that the field equation (\ref{FE1}) is differ from that of \cite%
{Yixin/2019} due to the choice of the sign in defining the non-metricity
scalar (\ref{Q}), yet it does not affect the results obtained in general. In
the present paper a perfect fluid type spacetime is considered, for which
the stress energy tensor is given by 
\begin{equation*}
T_{\mu \nu }=pg_{\mu \nu }+(p+\rho )u_{\mu }u_{\nu }\,,
\end{equation*}%
where $\rho $, $p$ and $u^{\mu }$ represent the energy density, pressure and
four velocity of the fluid respectively. We have chosen here the matter
Lagrangian to be $\mathcal{L}_{M}=p$. In addition, the matter Lagrangian is
supposed to rely only on the metric tensors. As a result, 
\begin{equation*}
\Theta _{\mu \nu }=pg_{\mu \nu }-2T_{\mu \nu }\,.
\end{equation*}

%%%%%%%%%%%%%%%%%%%%%%%%%%%%%%%%%%%%%%%%%%%%%%%%%%%%%%%%%%%%%%%%%
%%%%%%%%%%%%%%%%%%%%%%%%%%%%%%%%%%%%%%%%%%%%%%%%%%%%%%%%%%%%%%%%%

\section{Equations of motion in the LRS-BI model}

\label{sec3}

In this section, we consider the LRS-BI spacetime whose line element is
given by (\ref{metric}) in Cartesian coordinates. We use the usual flat
affine connection in this coincident gauge choice to obtain the motion
equations of a test particle. Corresponding to (\ref{metric}), the
directional Hubble parameters are defined as 
\begin{equation}
H_{x}\left( t\right) =\frac{\dot{A}}{A},\quad H_{y}\left( t\right) =\frac{%
\dot{B}}{B},\quad H_{z}\left( t\right) =\frac{\dot{B}}{B}\,,
\end{equation}%
and 
\begin{equation}
H\left( t\right) =\frac{1}{3}\frac{\dot{V}}{V}=\frac{1}{3}\left[ \frac{\dot{A%
}}{A}+2\frac{\dot{B}}{B}\right] \,,  \label{H}
\end{equation}%
is the average Hubble parameter, where the spatial volume is 
\begin{equation}
V\left( t\right) =AB^{2}=a^{3}.  \label{V}
\end{equation}%
where $a$ is the mean scale factor of the Universe.
The rate of expansion is evaluated by anisotropy parameter 
\begin{equation}
\Delta \left( t\right) =\frac{1}{3}\sum_{i=1}^{3}\left( \frac{H_{i}-H}{H}%
\right) ^{2}=\frac{2}{9H^{2}}\left( H_{x}-H_{y}\right) ^{2}.  \label{delta}
\end{equation}

It follows that 
\begin{equation}
H_{y}^{2}+2H_{x}H_{y}=3H^{2}\left( 1-\frac{\Delta }{2}\right) .  \label{36}
\end{equation}
The expansion scalar $\theta (t)$ and shear $\sigma (t)$ of the fluid are
given by 
\begin{equation}
\theta \left( t\right) =H_{x}+2H_{y},\quad \sigma \left( t\right) =\frac{|H_{x}-H_{y}|}{%
\sqrt{3}}.  \label{shear}
\end{equation}
The non-metricity scalar is given by 
\begin{equation}
Q\left( t\right) =-6(2H-H_{y})H_{y}\,.  \label{Q}
\end{equation}
Using \eqref{metric} and \eqref{FE1}, we obtain the following Friedmann-like
equations. 
\begin{align}
(\kappa +f_{T}\left( Q,T\right) )\rho +f_{T}\left( Q,T\right) p=& \frac{%
f\left( Q,T\right) }{2}+6f_{Q}\left( Q,T\right) (2H-H_{y})H_{y}\,,
\label{rho} \\
\kappa p=& -\frac{f\left( Q,T\right) }{2}-\frac{\partial }{\partial t}\left[
2f_{Q}\left( Q,T\right) H_{y}\right] -6f_{Q}\left( Q,T\right) H_{y}H,
\label{p_x} \\
\kappa p=& -\frac{f\left( Q,T\right) }{2}-\frac{\partial }{\partial t}\left[
f_{Q}\left( Q,T\right) (3H-H_{y})\right] -3f_{Q}\left( Q,T\right)
(3H-H_{y})H\,.  \label{p_y}
\end{align}
It follows from (\ref{rho})--(\ref{p_y}) that 
\begin{align}
\kappa \rho =& \frac{f\left( Q,T\right) }{2}+\frac{6f_{Q}\left( Q,T\right) }{%
\kappa +f_{T}\left( Q,T\right) }\left[ \kappa (2H-H_{y})H_{y}+f_{T}\left(
Q,T\right) H^{2}\right] +\frac{2f_{T}\left( Q,T\right) }{\kappa +f_{T}\left(
Q,T\right) }\frac{\partial }{\partial t}\left[ f_{Q}\left( Q,T\right) H%
\right] \,,  \label{rho-2} \\
\kappa p=& -\frac{f\left( Q,T\right) }{2}-2\frac{\partial }{\partial t}\left[
f_{Q}\left( Q,T\right) H\right] -6f_{Q}\left( Q,T\right) H^{2}\,.
\label{p-2}
\end{align}
On the other hand, using (\ref{p_x})--(\ref{p_y}) we obtain 
\begin{equation*}
0=\frac{\partial }{\partial t}\left[ f_{Q}\left( Q,T\right) (H-H_{y})\right]
+3f_{Q}\left( Q,T\right) (H-H_{y})H\,.
\end{equation*}
Solving this differential equation gives 
\begin{equation}
f_{Q}\left( Q,T\right) (H-H_{y})=la\left( t\right)^{-3}\,,  \label{DE}
\end{equation}%
where $l$ is a constant.

%%%%%%%%%%%%%%%%%%%%%%%%%%%%%%%%%%%%%%%%%%%%%%%%%%%%%%%%%%%%%%%%%
%%%%%%%%%%%%%%%%%%%%%%%%%%%%%%%%%%%%%%%%%%%%%%%%%%%%%%%%%%%%%%%%%

\section{Exact solutions of LRS-BI model}

\label{sec4}

In this section, we opt for the exact solutions of the above-mentioned
system which requires some additional assumption to determine. The
well-studied physical condition of proportionality between the expansion
scalar $\theta (t)$ and the shear scalar $\sigma (t)$ is used here, which
yields%
\begin{equation}
A\left( t\right) =B\left( t\right) ^{\lambda },  \label{ani}
\end{equation}%
where the constant $\lambda $ accounts for the anisotropic nature of the
model, i.e., if $\lambda $ is equal to one, the model is isotropic. The
physical basis for this hypothesis is supported by studies of the velocity
redshift relation for extragalactic sources, which indicate that the Hubble
expansion of the Universe can attain isotropy if $\frac{\sigma (t)}{\theta
(t)}$\ is constant. Collins proved the physical importance of this condition
in the case of a perfect fluid with a barotropic equation of state (EoS).
Additionally, it was showed \cite{adga} that in radiation era, a quadratic
model $f(Q)=f_{0}Q^{2}$ reproduces the condition $\sigma ^{2}\propto \theta
^{2}$.

Using the condition (\ref{ani}), we can derive the relationship between the
directional Hubble parameters $H_{x}\left( t\right) $ and $H_{y}\left(
t\right) $ as follows:%
\begin{equation}
H_{x}\left( t\right) =\lambda H_{y}\left( t\right) .  \label{Hx}
\end{equation}

Also, we obtain 
\begin{equation}
H_{y}\left( t\right) =\frac{3H\left( t\right) }{\lambda +2}\,.
\label{eq:H_y}
\end{equation}

Using Eqs. (\ref{DE}) and (\ref{eq:H_y}), we get 
\begin{equation}
f_{Q}\left( Q,T\right) H\left( t\right) =\frac{(\lambda +2)la\left( t\right)
^{-3}}{\left( \lambda -1\right) }\,,  \label{f_QH}
\end{equation}

{\ In this case, the non-metricity scalar }$Q\left( t\right) $ in Eq. (\ref%
{Q}){\ has the form%
\begin{equation}
Q\left( t\right) =-\frac{18(2\lambda +1)}{(\lambda +2)^{2}}H\left( t\right)
^{2}\,.  \label{Q-2}
\end{equation}%
} Substituting these into (\ref{rho-2})--(\ref{p-2}) and simplifying we get 
\begin{equation}
\kappa \rho ={\frac{f\left( Q,T\right) }{2}+\frac{6f_{Q}\left( Q,T\right)
H\left( t\right) ^{2}}{\kappa +f_{T}\left( Q,T\right) }\frac{3\kappa
(2\lambda +1)}{(\lambda +2)^{2}}}\,,  \label{F1}
\end{equation}%
\begin{equation}
\kappa p=-\frac{f\left( Q,T\right) }{2}\,.  \label{F2}
\end{equation}

From the previous two equations, we see that when studying the case of a
Universe filled with pressureless matter, i.e. $p=0$ in association with
condition (\ref{ani}), the function $f(Q,T)$ is not defined. Thus, in the
next two sections, we study some forms of the function $f\left( Q,T\right) $
with $p\neq 0$.

\subsection{Cosmological model with $f\left( Q,T\right) =\protect\alpha %
Q^{\left( n+1\right) }+\protect\beta T$}

\label{subsec1}

As a first case of the cosmological model of $f(Q,T)$ gravity, let us
consider the scenario when $f(Q,T)$ has the non-linear form $f\left(
Q,T\right) =\alpha Q^{\left( n+1\right) }+\beta T$, where $\alpha $, $\beta $
and $n$\ are constants. Then, we have $f_{Q}=\alpha (n+1)Q^{n}$ and $%
f_{T}=\beta $. Thus, Eq. (\ref{f_QH}) becomes 
\begin{equation}
\alpha (n+1)\left( \frac{-18(2\lambda +1)}{(\lambda +2)^{2}}\right)
^{n}H\left( t\right) ^{2n+1}=\frac{(\lambda +2)la\left( t\right) ^{-3}}{%
\left( \lambda -1\right) }\,.  \label{HQn}
\end{equation}

Solving this equation gives 
\begin{equation}
H(t)=\frac{2n+1}{3t+(2n+1)c_{n}}\,,\quad  \label{eqn:H:n}
\end{equation}%
where $c_{n}$ is a constant.

Again, solving the field equations (\ref{F1}) and (\ref{F2}) to find the
expressions for pressure and energy density, we obtain%
\begin{equation}
\rho \left( t\right) =-\frac{\alpha 2^{n-1} 9^{n+1} (\beta (3 n+2)+8 \pi (2
n+1)) \left(-\frac{H^2 (2 \lambda +1)}{(\lambda +2)^2}\right)^{n+1}}{(\beta
+4 \pi ) (\beta +8 \pi )},  \label{rho2}
\end{equation}%
\begin{equation}
p\left( t\right) =-\frac{\alpha 2^{n-1} 9^{n+1} (\beta (n+2)+8 \pi ) \left(-%
\frac{H^2 (2 \lambda +1)}{(\lambda +2)^2}\right)^{n+1}}{(\beta +4 \pi )
(\beta +8 \pi )}.  \label{p2}
\end{equation}

Therefore, at the first epoch $t\rightarrow0$, we see that energy density
and pressure have finite values. Also, these quantities diminish in value as
cosmic time $t$ increases and approaches to zero at infinite time.

Now, using Eqs. (\ref{Hx}) and (\ref{eq:H_y}), the directional Hubble
parameters are obtained as%
\begin{equation}
H_{x}\left( t\right) =\frac{3\lambda (2n+1)}{(\lambda +2)[3t+(2n+1)c_{n}]},
\end{equation}%
and%
\begin{equation}
H_{y}\left( t\right) =\frac{3(2n+1)}{(\lambda +2)[3t+(2n+1)c_{n}]}.
\end{equation}

Further, the metric potentials are derived as%
\begin{equation}
A\left( t\right) =\left( \frac{H_{0}[3t+(2n+1)c_{n}]}{2n+1}\right) ^{\frac{%
(2n+1)\lambda }{\lambda +2}},  \label{A2}
\end{equation}%
and%
\begin{equation}
B\left( t\right) =\left( \frac{H_{0}[3t+(2n+1)c_{n}]}{2n+1}\right) ^{\frac{%
2n+1}{\lambda +2}},  \label{B2}
\end{equation}%
where 
\begin{equation}
H_{0}=(\lambda +2)\left[ \frac{l}{(\lambda -1)(n+1)\alpha }\right] ^{\frac{1%
}{2n+1}}\left[ \frac{-1}{18(2\lambda +1)}\right] ^{\frac{n}{2n+1}}\,.
\end{equation}

The model exhibits no singularity at the beginning epoch $t\rightarrow 0$,
since the metric potentials have constant values. For time $t\rightarrow
\infty $, the metric potentials tend to infinity as time passes. Thus, using
Eqs. (\ref{A2}) and (\ref{B2}), the LRS Bianchi type-I metric becomes 
\begin{equation}
ds^{2}=-dt^{2}+\left( \frac{H_{0}[3t+(2n+1)c_{n}]}{2n+1}\right) ^{\frac{%
(2n+1)\lambda }{\lambda +2}}dx^{2}+\left( \frac{H_{0}[3t+(2n+1)c_{n}]}{2n+1}%
\right) ^{\frac{2n+1}{\lambda +2}}(dy^{2}+dz^{2}).
\end{equation}

By applying (\ref{V}), (\ref{shear}) and (\ref{Hx}), the spatial volume $V$,
expansion scalar $\theta (t)$, and shear scalar $\sigma (t)$ are derived as,%
\begin{equation}
V\left( t\right) =\left( \frac{H_{0}[3t+(2n+1)c_{n}]}{2n+1}\right) ^{2n+1},
\label{eqn:V2}
\end{equation}%
\begin{equation}
\theta (t)=\frac{3(2n+1)}{3t+(2n+1)c_{n}},  \label{eqn:theta2}
\end{equation}%
\begin{equation}
\sigma (t)=\frac{\sqrt{3}(\lambda -1)(2n+1)}{(\lambda +2)[3t+(2n+1)c_{n}]}.
\label{eqn:sigma2}
\end{equation}

Using Eqs. (\ref{eqn:theta2}) and (\ref{eqn:sigma2}), we obtain%
\begin{equation}
\frac{\sigma (t)^{2}}{\theta (t)}=\frac{(\lambda -1)^{2}(2n+1)}{(\lambda
+2)^{2}[3t+(2n+1)c_{n}]}.  \label{iso}
\end{equation}

Here, It is observed that the isotropy condition, i.e., $\frac{\sigma (t)^{2}%
}{\theta (t)}\rightarrow 0$ as $t\rightarrow \infty $, is fulfilled in this
case. {Eqs. (\ref{eqn:V2}) and (\ref{eqn:theta2}) show that the spatial
volume is finite at $t=0$ and increases with time from a finite to an
infinitely large value, but the expansion scalar is infinite, implying that
the Universe begins to evolve with finite volume at $t=0$. }This is
compatible with the Big Bang scenario.

Also, from Eq. (\ref{eqn:H:n}), we obtain%
\begin{equation}
\frac{H\left( t\right) }{H_{0}}=\frac{3t_{0}+(2n+1)c_{n}}{3t+(2n+1)c_{n}},
\label{H11}
\end{equation}%
where $H_{0}$ and $t_{0}$\ represent the current value of Hubble parameter
and age of the Universe. To get cosmological findings that enable for direct
comparison of cosmological model predictions with astronomical data, we use
the redshift parameter $z$ as an independent variable in place of the cosmic
time variable $t$. The redshift parameter for a distant source is inversely
related to the scale factor of the Universe at the moment in which the
photons were produced from the source. In this case, the equation relates
the scale factor $a\left( t\right) $ with the redshift parameter $z$ is
given by%
\begin{equation}
a=\frac{a_{0}}{\left( 1+z\right) }.  \label{az}
\end{equation}

Here, the scale factor is normalized such that its current value is one i.e. 
$a_{0}=a\left( 0\right) =1$. Using Eqs. (\ref{H11}) and (\ref{az}), the
expression for Hubble parameter $H\left( t\right) $ in terms of the redshift
parameter $z$ is derived as%
\begin{equation}
H\left( z\right) =H_{0}\left( 1+z\right) ^{\frac{3}{2n+1}},  \label{H(z)2}
\end{equation}%
where 
\begin{equation}
H_{0}=(\lambda +2)\left[ \frac{l}{(\lambda -1)(n+1)\alpha }\right] ^{\frac{1%
}{2n+1}}\left[ \frac{-1}{18(2\lambda +1)}\right] ^{\frac{n}{2n+1}}\,.
\end{equation}

In addition, we find the power-law expansion as the solution to the field
equations in an anisotropic Universe. Power-law cosmology provides an
attractive solution to several exceptional problems, including flatness and
the horizon problem. In literature, the power-law expansion is well
justified. The author of \cite{Kumar} examined cosmic parameters using a
power-law, and Hubble and Type Ia supernova datasets. Rani et al. \cite{Rani}
used state-finder analysis to investigate the power law cosmology. Recently,
Koussour and Bennai \cite{Kous} employed the power law to investigate cosmic
acceleration in an anisotropic Universe in $f(Q)$ gravity.

The deceleration parameter $q\left( t\right) $, which indicates the
accelerating/decelerating aspect of the expansion of the Universe, is an
essential cosmological variable. The deceleration parameter is expressed by
the equation:%
\begin{equation}
q=-1+\frac{d}{dt}\left( \frac{1}{H}\right) =-1-\frac{\overset{.}{H}}{H^{2}}.
\end{equation}

Using Eq. (\ref{az}), the time operator is given by%
\begin{equation}
\frac{d}{dt}=\frac{dz}{dt}\frac{d}{dz}=-\left( 1+z\right) H\left( z\right) 
\frac{d}{dz}.
\end{equation}

The deceleration parameter $q\left( t\right) $ can be calculated as a
function of the redshift parameter $z$ using the formula%
\begin{equation}
q\left( z\right) =-1+\left( 1+z\right) \frac{1}{H\left( z\right) }\frac{%
dH\left( z\right) }{dz}.
\end{equation}

Here, the sign of the $q\left( z\right) $ (negative or positive) shows if
the Universe accelerates or decelerates. The value of the deceleration
parameter for this scenario is 
\begin{equation}
q=-1+\frac{3}{1+2n}.  \label{qq}
\end{equation}

\begin{figure}[h]
\centerline{\includegraphics[scale=0.80]{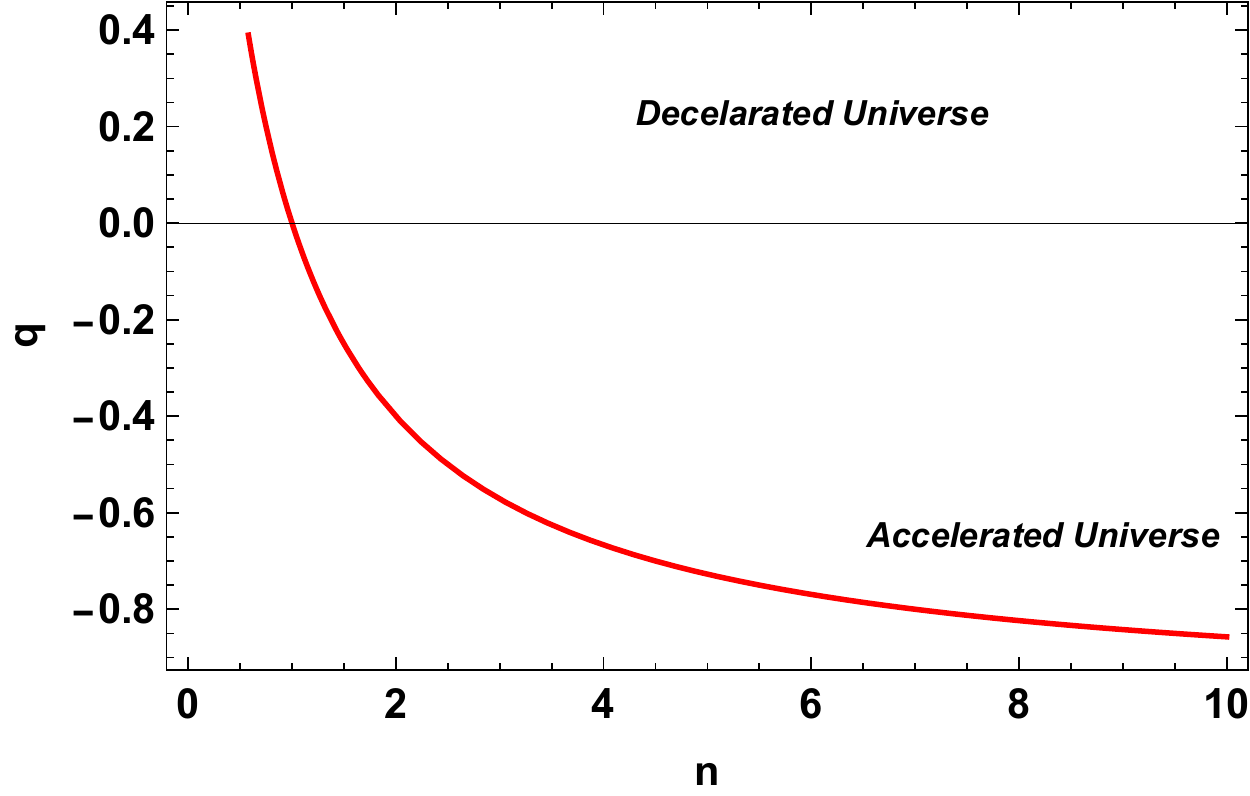}}
\caption{The plot of deceleration parameter vs. $n$.}
\label{q}
\end{figure}

This is a constant as predicted according to the power law type expansion of
the cosmological model. It is clear from Eq. (\ref{qq}) that there is a
transition phase from deceleration to acceleration at $n=1$. Fig. \ref{q}
depicts the evolution of the deceleration parameter in terms of $n$ for the
function $f\left( Q,T\right) =\alpha Q^{\left( n+1\right) }+\beta T$ and in
our model, it is directly dependent on the parameter $n$. According to Fig. %
\ref{q}, the deceleration parameter is positive at $n<1$ and negative for $%
n>1$. Thus, it shows that the Universe is transitioning from deceleration to
acceleration. $q$ decreases as $n$ increases.

\subsection{Cosmological model with $f\left( Q,T\right) =\protect\alpha Q+%
\protect\beta T$}

\label{subsec2}

We can observe that for $n=0$, the previous situation is reduced to the
linear form of the $f\left( Q,T\right) $ function as $f\left( Q,T\right)
=\alpha Q+\beta T$, where $\alpha $ and $\beta $ are constants. Then, we
have $f_{Q}=\alpha $ and $f_{T}=\beta $. So, for $n=0$, we can solve Eq. (%
\ref{HQn}) and get the Hubble parameter expression as%
\begin{equation}
H\left( t\right) =\frac{1}{3t+c_{0}},  \label{H1}
\end{equation}%
where $c_{0}$ is the constant i.e. $c_{0}=c_{n}\left( n=0\right) $.

Now, using Eqs. (\ref{Hx}) and (\ref{eq:H_y}), the directional Hubble
parameters are obtained as%
\begin{equation}
H_{x}\left( t\right) =\frac{3\lambda }{(\lambda +2)\left[ 3t+c_{0}\right] },
\end{equation}%
and%
\begin{equation}
H_{y}\left( t\right) =\frac{3}{(\lambda +2)\left[ 3t+c_{0}\right] }.
\end{equation}

Using Eq. (\ref{iso}), we obtain%
\begin{equation}
\frac{\sigma (t)^{2}}{\theta (t)}=\frac{(\lambda -1)^{2}}{(\lambda +2)^{2}%
\left[ 3t+c_{0}\right] }.
\end{equation}

Also, it is observed that the isotropy condition, i.e., $\frac{\sigma (t)^{2}%
}{\theta (t)}\rightarrow 0$ as $t\rightarrow \infty $, is fulfilled in this
case.

As previously mentioned, all of the cosmological parameters listed above
must be expressed in terms of redshift parameter $z$. Using Eq. (\ref{H(z)2}%
), the expression of Hubble parameter $H\left( t\right) $ for $n=0$ in terms
of the redshift parameter $z$ is derived as%
\begin{equation}
H\left( z\right) =H_{0}\left( 1+z\right) ^{3},  \label{H111}
\end{equation}%
where $H_{0}=\frac{(\lambda +2)l}{\alpha (\lambda -1)}$. It is important to
note that Eq. (\ref{H(z)2}) can be interpreted like the famous Hubble's Law
which states that the proper distance $d$ between galaxies is proportional
to their recessional velocity $v$ as measured by the Doppler effect redshift
i.e. $v=H_{0}d$. Thus, the value of the Hubble parameter in terms of
redshift parameter is extremely important in an astrophysical background.
For this scenario, the value of the deceleration parameter is $q=2$,
implying a Universe that is decelerating. Many authors of different modified
gravity theories have obtained the same result.

\section{Observational constraints}

\label{sec6}

It should be highlighted that a thorough evaluation of the parameter values
is important in examining the cosmological features. In this sense, the
current section presents observational analyses of the current situation.
The statistical method we use helps us to constrain parameters like $H_{0}$
and $n$. Especially, we use the Markov Chain Monte Carlo (MCMC) with the
standard Bayesian approach. Furthermore, using the pseudo-chi-squared
function $\chi ^{2}$, the best fit values for the parameters are obtained by
the probability function,%
\begin{equation}
\mathcal{L}\propto e^{-\frac{\chi ^{2}}{2}},
\end{equation}

To do this, we now focus on two datasets: Hubble and Type Ia supernova (SNe
Ia) data. To begin, we evaluate the parameter space priors, which are $%
(60.0<H_{0}<80.0)$ to account for all possible scenarios of the Hubble
parameter, $(-10.0<n<+10.0)$ to get all the scenarios of the expansion of
the Universe. Also, take into consideration that our cosmological model must
also fit the observational datasets. The next subsections go into further
depth on the data sets and statistical analyses.

\subsection{Hubble datasets}

Here, we employ a standardized collection of $31$ measures derived from the
differential age technique (DA) in the redshift range $0.07<z<2.42$ and are
listed in Tab. \ref{tab1} \cite{Moresco,Ratra}. The DA technique can be used
to calculate the rate of expansion of the Universe at redshifts $z$. The
Hubble datasets chi-square ($\chi ^{2}$) is calculated as follows: 
\begin{equation}
\chi _{Hubble}^{2}=\sum_{j=1}^{31}\frac{\left[
H_{th}(z_{j})-H_{obs}(z_{j},p_{s})\right] ^{2}}{\sigma (z_{j})^{2}},
\end{equation}%
where $H_{th}$ and $H_{obs}$ are the theoretical and observed values of the
Hubble parameter $H\left( z\right) $, and $p_{s}$ denotes the parameter
space of the model to be constrained. In addition, $\sigma ^{2}$ denotes the
standard error in the observed value of $H\left( z\right) $, $p_{s}=\left(
H_{0},n\right) $ and is the parameter space of the cosmic background. 
\begin{table}[h]
\label{tab1}%
\begin{tabular}{||c|c|c|c||c|c|c|c||}
\hline
$z$ & $H(z)$ & $\sigma_H$ & Ref. & $z$ & $H(z)$ & $\sigma_H$ & Ref. \\%
[0.5ex] \hline\hline
$0.070$ & $69$ & $19.6$ & \cite{Stern/2010} & $0.4783$ & $80$ & $99$ & \cite%
{Moresco/2016} \\ \hline
$0.90$ & $69$ & $12$ & \cite{Simon/2005} & $0.480$ & $97$ & $62$ & \cite%
{Stern/2010} \\ \hline
$0.120$ & $68.6$ & $26.2$ & \cite{Stern/2010} & $0.593$ & $104$ & $13$ & 
\cite{Moresco/2012} \\ \hline
$0.170$ & $83$ & $8$ & \cite{Simon/2005} & $0.6797$ & $92$ & $8$ & \cite%
{Moresco/2012} \\ \hline
$0.1791$ & $75$ & $4$ & \cite{Moresco/2012} & $0.7812$ & $105$ & $12$ & \cite%
{Moresco/2012} \\ \hline
$0.1993$ & $75$ & $5$ & \cite{Moresco/2012} & $0.8754$ & $125$ & $17$ & \cite%
{Moresco/2012} \\ \hline
$0.200$ & $72.9$ & $29.6$ & \cite{Zhang/2014} & $0.880$ & $90$ & $40$ & \cite%
{Stern/2010} \\ \hline
$0.270$ & $77$ & $14$ & \cite{Simon/2005} & $0.900$ & $117$ & $23$ & \cite%
{Simon/2005} \\ \hline
$0.280$ & $88.8$ & $36.6$ & \cite{Zhang/2014} & $1.037$ & $154$ & $20$ & 
\cite{Moresco/2012} \\ \hline
$0.3519$ & $83$ & $14$ & \cite{Moresco/2012} & $1.300$ & $168$ & $17$ & \cite%
{Simon/2005} \\ \hline
$0.3802$ & $83$ & $13.5$ & \cite{Moresco/2016} & $1.363$ & $160$ & $33.6$ & 
\cite{Moresco/2015} \\ \hline
$0.400$ & $95$ & $17$ & \cite{Simon/2005} & $1.430$ & $177$ & $18$ & \cite%
{Simon/2005} \\ \hline
$0.4004$ & $77$ & $10.2$ & \cite{Moresco/2016} & $1.530$ & $140$ & $14$ & 
\cite{Simon/2005} \\ \hline
$0.4247$ & $87.1$ & $11.2$ & \cite{Moresco/2016} & $1.750$ & $202$ & $40$ & 
\cite{Simon/2005} \\ \hline
$0.4497$ & $92.8$ & $12.9$ & \cite{Moresco/2016} & $1.965$ & $186.5$ & $50.4$
& \cite{Moresco/2015} \\ \hline
$0.470$ & $89$ & $34$ & \cite{Ratsimbazafy/2017} &  &  &  &  \\ \hline
\end{tabular}%
\caption{$H(z)$ datasets with 31 data points.}
\end{table}
\begin{figure}[h]
\centerline{\includegraphics[scale=0.60]{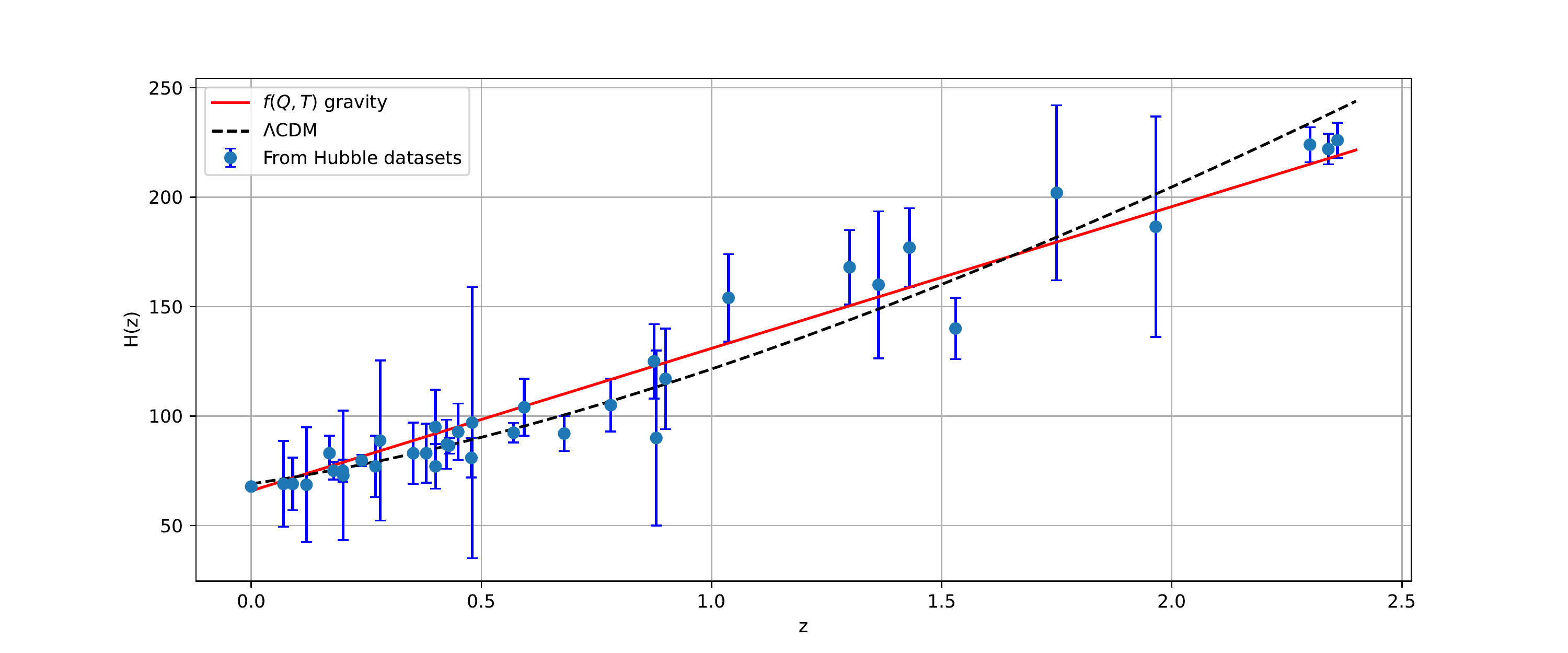}}
\caption{The plot of $H(z)$ vs. redshift parameter $z$ for our $f(Q,T)=%
\protect\alpha Q^{\left( n+1\right) }+\protect\beta T$ model, shown in red,
and $\Lambda $CDM, shown in black dashed lines, shows an excellent fit to
the 31 points of the Hubble datasets.}
\label{ErrorHubble}
\end{figure}

\subsection{Type Ia supernovae (SNe Ia) datasets}

The measurement of SNe Ia is essential to comprehend how the Universe is
expanding. The Panoramic Survey Telescope and Rapid Response System
(Pan-STARSS1), Sloan Digital Sky Survey (SDSS), Supernova Legacy Survey
(SNLS), and Hubble Space Telescope (HST) surveys all collected data on SNe
Ia \cite{Scolnic}. Here, we employ the Pantheon sample, which consists of
1048 points with distance moduli $\mu _{j}$ in the range $0.01<z_{j}<2.26$
at various redshifts. The SNe Ia datasets chi-square ($\chi ^{2}$) is
calculated as follows:%
\begin{equation}
\chi _{SNe}^{2}=\sum_{j,i=1}^{1048}\Delta \mu _{j}(C_{SNe}^{-1})_{ji}\Delta
\mu _{i}.
\end{equation}

Here, $C_{SNe}$ is the covariance matrix, and $\Delta \mu _{j}=\mu
_{j}^{th}(z_{j},p_{s})-\mu _{j}^{obs}$ is the difference between the
measured distance modulus value collected from cosmic measurements and its
theoretical values estimated from the model with the specified parameter
space $p_{s}$. The theoretical distance modulus $\mu ^{th}$ is given as 
\begin{equation}
\mu ^{th}(z_{j})=25+5log_{10}\left[ \frac{d_{l}}{1Mpc}\right] ,
\end{equation}%
and the luminosity distance $d_{l}$ defined as,%
\begin{equation}
d_{l}(z)=(1+z)\int_{0}^{z}\frac{c}{H(z,p_{s})}.
\end{equation}

\begin{figure}[h]
\centerline{\includegraphics[scale=0.60]{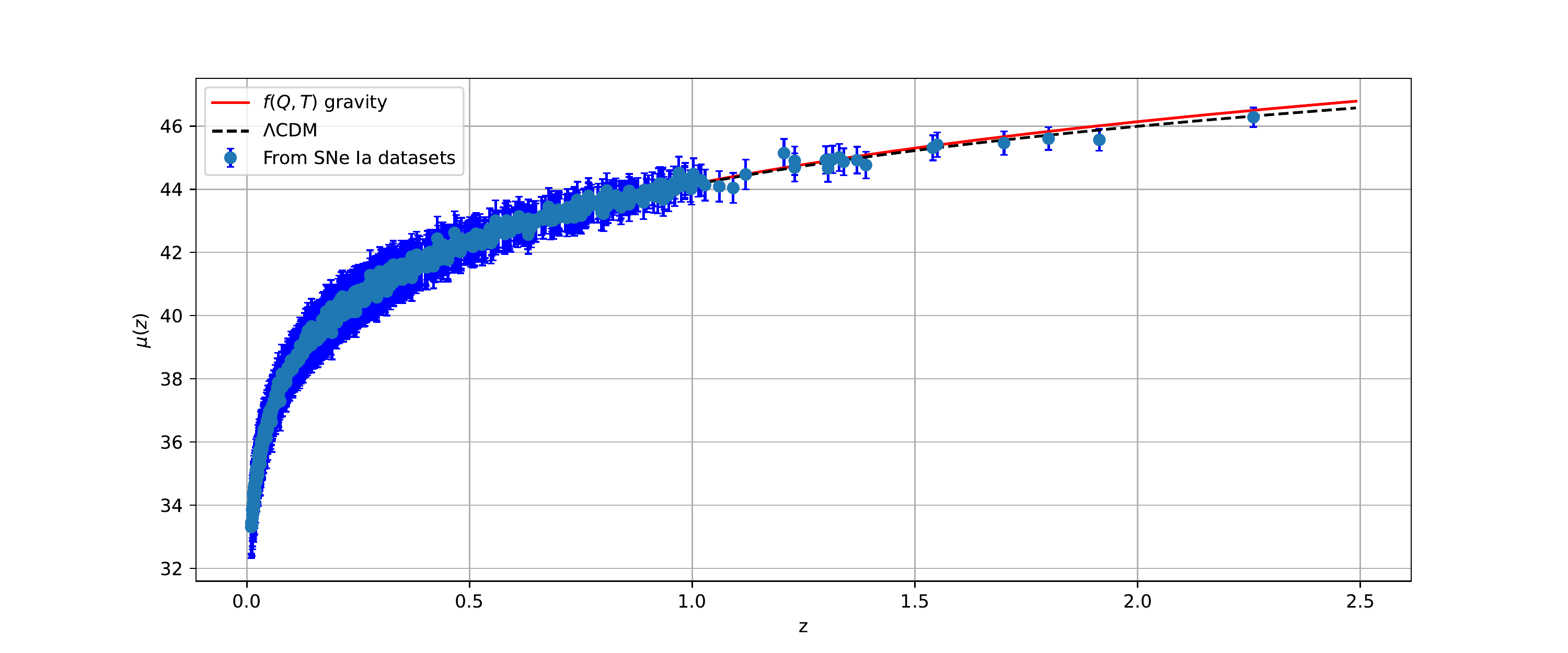}}
\caption{The plot of $\protect\mu (z)$ vs. redshift parameter $z$ for our $%
f(Q,T)=\protect\alpha Q^{\left( n+1\right) }+\protect\beta T$ model, shown
in red, and $\Lambda $CDM, shown in black dashed lines, shows an excellent
fit to the 1048 points of the SNe Ia datasets.}
\label{ErrorMu}
\end{figure}

\subsection{Results}

\begin{figure*}[th]
\subfloat[\label{genworkflow}]{      \includegraphics[trim=1 4 10 20,clip,
width=0.3\textwidth]{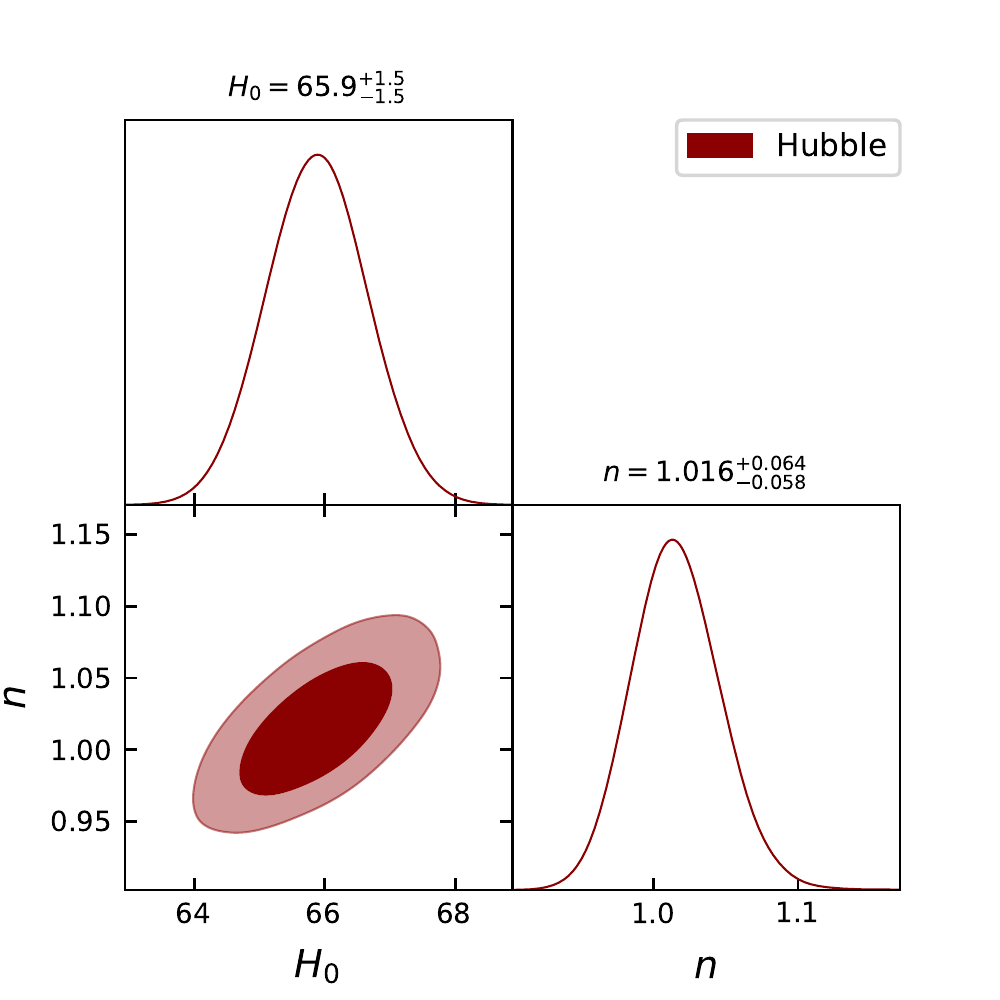}} \hspace{\fill} 
\subfloat[\label{pyramidprocess} ]{      \includegraphics[trim=1 4 10 20,clip,
width=0.3\textwidth]{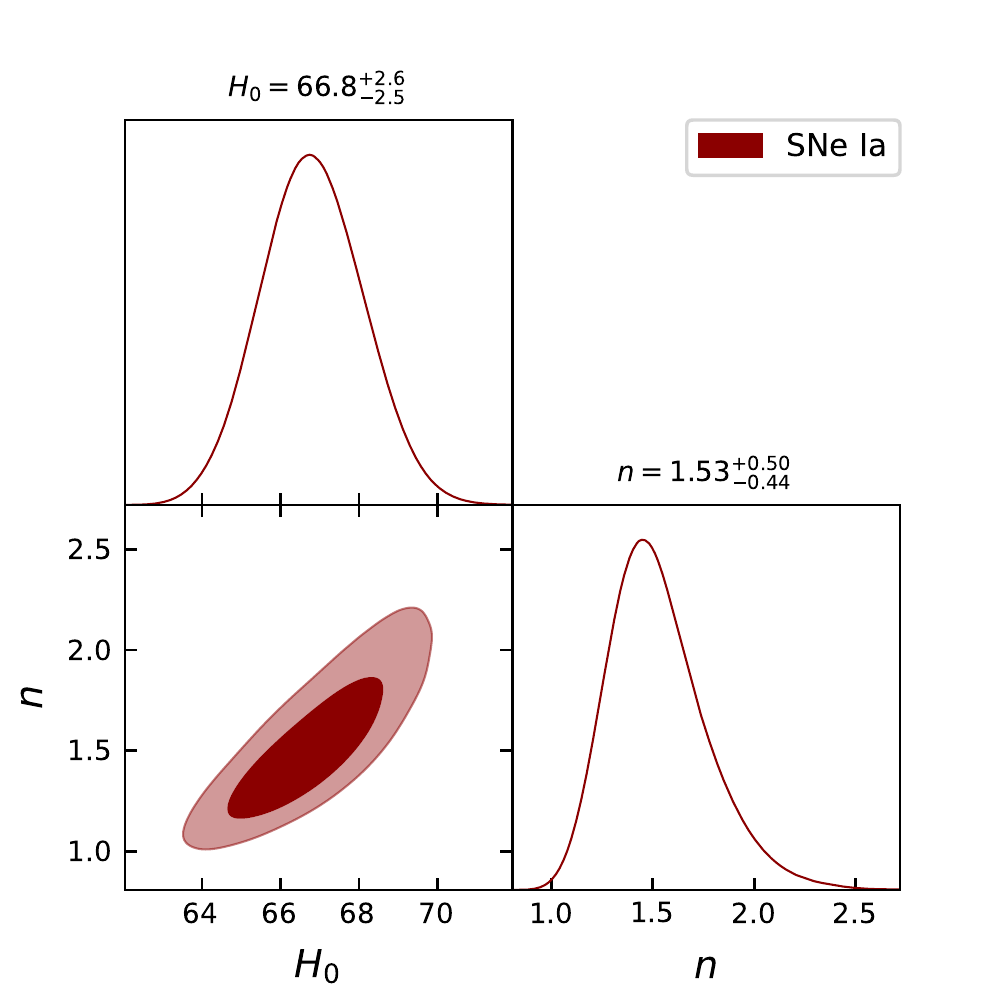}} \hspace{\fill} 
\subfloat[\label{mt-simtask}]{      \includegraphics[trim=1 4 10 20,clip,
width=0.3\textwidth]{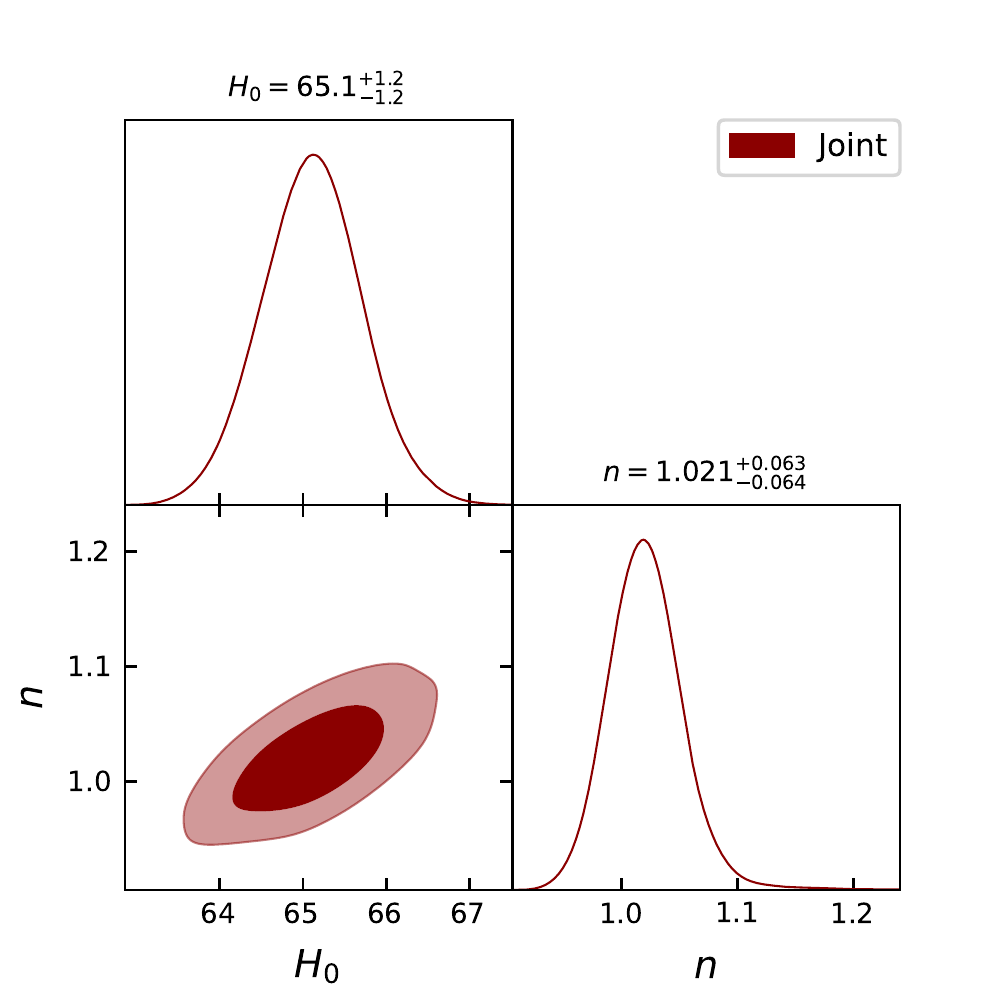}}\newline
\caption{The marginalized constraints on the parameters $H_{0}$ and $n$ are
presented using the Hubble (a), SNe Ia (b), and Joint (c) datasets. The dark red shaded
areas represent the $1-\protect\sigma $ confidence level (CL), whereas the
light red shaded regions represent the $2-\protect\sigma $ confidence level.
The parameter constraint values are displayed at the $1-\protect\sigma $ CL.}
\label{Con}
\end{figure*}

The model parameters for the joint (Hubble+SNe) are constrained using $\chi
^{2}=\chi _{Hubble}^{2}+\chi _{SNe}^{2}$. Tab. \ref{tab} shows the outcomes
and results. Figs. \ref{ErrorHubble} and \ref{ErrorMu} compare our model to
the widely accepted $\Lambda $CDM model in cosmology i.e. $H\left( z\right) =H_{0}
\sqrt{\Omega
_{0}^{m}\left( 1+z\right) ^{3}+\Omega _{0}^{\Lambda }}$; we use $\Omega
_{0}^{m}=0.3$, $\Omega _{0}^{\Lambda }=0.7$, and $H_{0}=69$ $%
km.s^{-1}.Mpc^{-1}$ for the plot. The figures also show the Hubble and SNe
Ia experimental findings, with 31 and 1048 data points and errors,
respectively, allowing for a direct comparison of the two models. To find
out the outcomes of our MCMC study, we employed 100 walkers and 1000 steps
for all datasets: Hubble, SNe Ia, and Joint. Also, Fig. \ref{Con} shows the
likelihood contours for Hubble, SNe Ia, and Joint analysis, and Tab. \ref%
{tab} shows the numerical findings. From Fig. \ref{Con}, it is clear that
the likelihood functions for all datasets (Hubble, SNe Ia, and Joint) are
very well matched to a Gaussian distribution function. In every cosmological
model, the Hubble constant $H_{0}$ and the deceleration parameter $q$ play a
significant role in characterizing the nature of the expansion of the
Universe. The first describes the current rate of expansion of the Universe,
whereas the latter describes if the Universe is accelerating $(q<0)$ or
decelerating $(q>0)$. We obtain the constraints on these parameters using
the most recent Hubble with 31 data points and SNe Ia data with 1048
pantheon sample points. At the $1-\sigma $ CL, the constraints determined
from Hubble datasets are $H_{0}=65.9^{+1.5}_{-1.5}$ and $q=-0.011\pm 0.01$,
whereas the constraints determined from SNe Ia data are $%
H_{0}=66.8^{+2.6}_{-2.5}$ and $q=-0.261\pm0.03$. We also run the joint test
with Hubble and SNe Ia datasets, which gives the constraints $%
H_{0}=65.1^{+1.2}_{-1.2}$ and $q=-0.014\pm 0.01$. It is worth noting that
the values of parameter $H_{0}$ correspond to the observations \cite%
{Planck2020}. Also, the deceleration parameter values show that the
observational data represent the actual cosmic acceleration within the
context of anisotropic $f(Q,T)$ cosmology.

\begin{table*}[th]
\caption{With a confidence level of 68\%, marginalized constrained data of
the parameters $H_{0}$, $n$ and $q$ for various data samples were obtained.}
\label{tab}%
\begin{ruledtabular}
		    \centering
		    \begin{tabular}{c c c c c}
				Dataset & $H_{0}$ & $n$ & $q$\\ 
				\hline
				$Hubble$ & $65.9^{+1.5}_{-1.5}$ & $1.016^{+0.064}_{-0.058}$ & $-0.011\pm 0.01$\\
				$SNe Ia$ & $66.8^{+2.6}_{-2.5}$  & $1.53^{+0.50}_{-0.44}$& $-0.261\pm0.03$\\
				$Joint$ & $65.1^{+1.2}_{-1.2}$ & $1.021^{+0.063}_{-0.046}$ & $-0.014\pm 0.01$\\
			\end{tabular}
		   \end{ruledtabular}
\end{table*}

\section{Concluding remarks}

\label{conclusion}

This study has investigated an anisotropic cosmology in the modified $f(Q,T)$
gravity theory, where $Q$ denotes non-metricity scalar and $T$ is the trace
of the energy-momentum tensor. The exact solution of the field equations for
an LRS Bianchi type-I spacetime are explored. Because the field equations
are highly nonlinear and difficult, we have solved them by assuming that the
expansion scalar $\theta (t)$ is proportional to the shear scalar $\sigma (t)
$. It has provided $A\left( t\right) =B\left( t\right) ^{\lambda }$, where $%
A(t)$ and $B(t)$ are the metric potentials and $\lambda$ is an arbitrary
constant that accounts for the anisotropic nature of the model. We have
primarily investigated two solutions of modified field equations using two
functional forms of $f(Q,T)$.

For the model $f(Q,T)=\alpha Q^{n+1}+\beta T$ the isotropy condition, i.e. $%
\frac{\sigma (t)^{2}}{\theta (t)}\rightarrow 0$ as $t\rightarrow \infty $,
has been fulfilled. At $t=0$, the spatial volume is finite and the expansion
scalar is infinite, implying that the Universe began to evolve with a finite
volume at $t = 0$. We obtain the scenario of the big bang. The energy
density and pressure are finite at the first epoch. Furthermore, as cosmic
time $t$ increases, the value of these quantities decreases and approaches $0
$ at infinite time. The deceleration parameter $q(t)$ is found to be $q=-1+%
\frac{3}{1+2n}$, implying a phase transition from deceleration to
acceleration at $n=1$. From Fig. \ref{q}, it is observed that the
deceleration parameter is positive (deceleration) at $n<1$ and negative
(acceleration) for $n>1$. Next, to obtain the constraint value for the
parameter $n$, we employed the statistical Markov chain Monte Carlo (MCMC)
method with the Bayesian approach. We also examined the results for two
independent observational datasets, Hubble datasets, and Type Ia supernovae
(SNe Ia) datasets, which contain SDSS, SNLS, Pan-STARRS1, low-redshift
survey, and HST surveys. The best-fit values obtained are $%
n=1.016^{+0.064}_{-0.058}$ for the Hubble datasets, $n=1.53^{+0.50}_{-0.44}$
for the SNe Ia datasets and $n=1.021^{+0.063}_{-0.046}$ for the Hubble+SNe
Ia datasets. Moreover, the deceleration parameter has been constrained,
which is important in describing the evolution of the Universe. The best-fit
values obtained are $q=-0.011\pm 0.01$ for the Hubble datasets, $%
q=-0.261\pm0.03$ for the SN Ia datasets and $q=-0.014\pm 0.01$ for the
Hubble+SNe Ia datasets, which indicates an accelerating model of the
Universe. Furthermore, using these parameter values, we compared our $%
f(Q,T)=\alpha Q^{n+1}+\beta T$ model with the most commonly accepted model
for the Universe i.e. $\Lambda$CDM in Figs. \ref{ErrorHubble} and \ref%
{ErrorMu}.

Finally, we have obtained the solutions to the field equation by studying
the linear case of the function $f(Q,T)$ i.e. $n=0$. We have observed the
same behavior for the cosmological parameters mentioned above, with the
exception that the deceleration parameter in this scenario turns out to be a
constant $q=2$, which indicates a decelerating model of the Universe. The paper by Shamir \cite{B2} discusses a cosmological model based on $f(R,T)$ gravity in LRS-BI space-time. The author derived the field equations and solved them using an analytical approach. The obtained solutions were then used to investigate the evolution of the scale factor, energy density, and pressure of the universe. When compared to our paper on $f(Q,T)$ gravity, the primary distinction between Shamir's study on LRS-BI cosmology in $f(R,T)$ gravity is that while our paper utilizes observational constraints on the parameters of the $f(Q,T)$ gravity model, especially, we used the MCMC method to fit our $f(Q,T)$ gravity model parameters to Hubble and SNe Ia datasets which allows us to estimate the posterior probability distribution of the model parameters and quantify the uncertainties in our parameter estimates, Shamir does not incorporate any observational data in their analysis of the $f(R,T)$ gravity model. By incorporating observational data in our analysis, our paper provides a more comprehensive study of the $f(Q,T)$ gravity model, as it allows for a more rigorous comparison between the theoretical predictions and observational data. This can lead to more robust constraints on the model parameters, which can help to rule out or support specific modifications to gravity theory. In addition, Kennedy et al. \cite{Kennedy} have reconstructed Horndeski's theories from phenomenological modified gravity and dark energy models on cosmological scales. Their approach is complementary to ours, as they aim to reconstruct the theoretical framework of Horndeski gravity from observed data, rather than proposing a specific gravity model. However, it is interesting to compare their results with ours, as both approaches aim to explain the observed acceleration of the universe without introducing dark energy.

Thus, these findings can encourage us to investigate the anisotropic nature
of $f (Q,T)$ theory further, since it follows the observational data.
Furthermore, it would be interesting to study the acceleration scenario of
the Universe using certain parameterizations of the equation of state
parameters. We intend to investigate this scenario in the future.

\section*{Data Availability}

All generated data are included in this manuscript.

%%%%%%%%%%%%%%%%%%%%%%%%%%%%%%%%%%%%%%%%%%%%%%%%%%%%%%%%%%%%%%%%%%%%%%%%%%%%%%%

\end{document}